\documentclass[article,reqno]{amsart}
\usepackage{amsmath,amssymb}
\usepackage{latexsym}
\usepackage{epsfig}
\usepackage{cite}
\usepackage{booktabs}
\usepackage{color}
\usepackage{graphicx}
\usepackage{tikz}
\usepackage{pgfplots}
\usepackage{amsfonts}
\usepackage{amsmath}

\usepackage{amsthm}
\newcommand{\beq}{\begin{equation}}
\newcommand{\eeq}{\end{equation}}
\newcommand{\beqn}{\begin{eqnarray}}
\newcommand{\eeqn}{\end{eqnarray}}
\newcommand{\chimag}{\chi_>}
\newcommand{\chimin}{\chi_<}
\newcommand{\chiI}{\chi_I}

\numberwithin{equation}{section}

\title{Spectral boundary conditions and solitonic solutions in a classical Sellmeier dielectric}
\author{F. Belgiorno$^{1,2,4}$, S.L. Cacciatori$^{3,4}$ \and A. Vigan\`o}
\address{\noindent $^1$Dipartimento di Matematica, Politecnico di Milano, Piazza Leonardo 32, IT-20133 Milano, Italy\endgraf
$^2$INdAM-GNFM \endgraf
$^3$Department of Science and High Technology, Universit\`a dell'Insubria, Via Valleggio 11, IT-22100 Como, Italy\endgraf
$^4$INFN sezione di Milano, via Celoria 16, IT-20133 Milano, Italy}

\begin{document}
\maketitle
\begin{abstract}
Electromagnetic field interactions in a dielectric medium represent a longstanding field of investigation, both at the 
classical level and at the quantum one. We propose a 1+1 dimensional toy-model which consists of an half-line filling dielectric medium, with the aim to set up a simplified situation where technicalities related to gauge invariance and, as a consequence, 
physics of constrained systems are avoided, and still interesting features appear. In particular, we simulate the electromagnetic field and the polarization field by means of two coupled scalar fields $\phi$,$\psi$ respectively, in a Hopfield-like model. 
We find that, in order to obtain a physically meaningful behaviour for the model, one has to introduce spectral boundary conditions depending on the particle spectrum one is dealing with. This is the first interesting achievement of our analysis. The second relevant achievement is that, by introducing a nonlinear contribution in the polarization field $\psi$, with the aim of mimicking a third order nonlinearity in a nonlinear dielectric, we obtain solitonic solutions in the Hopfield model framework, whose classical behaviour is 
analyzed too.  
\end{abstract}


\section{introduction}

In the framework of electromagnetic field interactions in a dielectric medium, both at the 
classical level and at the quantum one, a very rich phenomenology appears, involving several phenomena, from standard dispersion law to Hawking-like pair creation. 
We have developed in our previous studies an analysis of the Hopfield model, which has be made relativistically covariant, and suitably extended in order to keep into account in a semi-phenomenological way the possibility that 
e.g. the dielectric susceptibility (and/or the resonance frequency) depends on spacetime variables, with the aim of simulate the 
standard Kerr effect in nonlinear dielectric \cite{physicascripta,hopfield-hawking}. Quantization has been taken into account 
in \cite{physicascripta,annals}. The analysis with scalar models has been developed in \cite{hopfield-hawking,fipsi}, with the 
aim of gain knowledge of the basic physics at hand without all tricky technicalities which are associated with gauge invariance. 
A further step towards a more complete analysis is contained in \cite{nonlinear-hawking}, where a full four dimensional 
electromagnetic field in a nonlinear dielectric medium and the analogue Hawking effect have been investigated.\\
As a further contribution to our investigation of the Hopfield model, we extend our analysis by considering a dielectric medium 
which does not fill all the space as in our previous works. 
We propose a 1+1 dimensional toy-model which consists of an half-line filling dielectric medium, with the aim to set up a simplified situation where technicalities related to gauge invariance and, as a consequence, 
physics of constrained systems are avoided, and still interesting features appear. In particular, we simulate the electromagnetic field and the polarization field by means of two coupled scalar fields which are indicated as $\phi$,$\psi$ respectively, in a model which is inherited by the Hopfield model. The interface between the vacuum region and the dielectric one is represented 
by $z=0$, and the dielectric medium fills the region $z\geq 0$. 
The electromagnetic field $\phi$ is involved with both the vacuum region and the dielectric one, whereas the polarization 
$\psi$ is different from zero only in the dielectric region. By analyzing the particle spectrum of the model we find that, 
in order to obtain a physically meaningful behaviour, one has to introduce boundary conditions depending on the particle spectrum one is dealing with. Indeed, for the electromagnetic field one finds that smooth solutions with continuous $\phi,\partial_z \phi$ at the 
interface $z=0$ does not correspond to a complete scattering basis, due to the presence of a spectral gap (i.e. a gap in the 
particle spectrum) associated with the presence of the dielectric medium. Note that, for simplicity, we are purposefully 
dealing with transparent dielectric medium (absorption would require further efforts at the quantum level). For particles in the 
spectral gap, we have to impose Dirichlet boundary conditions at the interface, meaning a complete reflection for the associated electromagnetic modes.
This is the first interesting achievement of our analysis.\\ 
We also introduce a nonlinear contribution in the polarization field $\psi$, with the aim of mimicking a third order nonlinearity in a nonlinear dielectric medium.  
We obtain exact solutions, which correspond to propagating solitons, and study their energy propagation both in a global sense (spatial integrals) and in a local one (Poynting vectors). This is the second achievement of our analysis.\\
It is worth mentioning that there exists a huge literature concerning electromagnetic field 
in presence of a dielectric medium filling an half space, mainly in a framework where phenomenological refractive index appears. 
We limit to quote a classical textbook \cite{born-wolf}, for classical scattering of light, and 
the seminal study \cite{maradudin}, concerning the effects of spatial dispersion. See also \cite{bishop} on energy propagation. As to quantization 
of the 1+1 dimensional system, we refer to \cite{kireev,santos}, where a phenomenological approach to the electromagnetic field in inhomogeneous and dispersive media is assumed.  

\section{The half-line filling model}
We will consider a $1+1$ dimensional problem, where a straight line, parametrized by the coordinate $z$, is filled by a dielectric medium for $z\geq0$. The dielectric is described by a field $\psi$, and interacts with a ``scalar'' electromagnetic field $\phi$.
The system is described by the action
\begin{eqnarray}
\label{action1}
S[\phi,\psi]= \int_{\mathbb R} dt \left[ \int_{\mathbb R} \frac 12 \partial_\mu \phi \partial^\mu \phi \ dz +\int_{z\geq 0} [\frac 12 \dot \psi^2- \frac {\omega_0^2}2 \psi^2-g\phi\dot\psi ]dz  \right],
\end{eqnarray}
where the dot indicates time derivative.\\
We require for $\psi$ to be smooth in $z\geq 0$, to vanish elsewhere, but we do not add continuity conditions in $z=0$. For $\phi$ we require to be smooth in $z\geq 0$, and in $z<0$, and to be of class $C^1(\mathbb R)$. Our aim is to show that such 
boundary conditions are not sufficient in order to get a ``good problem''.\\
The equations of motion are thus
\begin{align}
&\Box \phi=0, & \psi=0, && \mbox{for } z<0, \\
\label{eqdiel}
&\Box \phi+g\dot\psi=0, & \ddot \psi+\omega_0^2 \psi-g\dot \phi=0, && \mbox{for } z\geq0,
\end{align}
with the condition
\begin{eqnarray}
\phi(0^+)=\phi (0^-), \qquad\ \phi'(0^+)=\phi' (0^-),
\end{eqnarray}
where the prime indicates spatial derivative.\\
We further require for the energy to be finite, which is equivalent to the condition
\begin{eqnarray}
\int_{\mathbb R} [\dot \phi^2+\phi^{\prime 2}+\dot\psi^2+\omega_0^2\psi^2]dz<\infty.\label{finite-energy}
\end{eqnarray}
Physically, any initial condition compatible with (\ref{finite-energy}) should be possible.
\subsection{General solution and plane wave bases}
We can take the Fourier transform of the fields in order to get the general solution. If we define $\phi=\phi_<+\phi_\geq$, where
\begin{eqnarray}
&& \phi_<(t,z)=\phi(t,z)\chi_{(-\infty,0)}(z),\\
&& \phi_\geq(t,z)=\phi(t,z)\chi_{[0,\infty)}(z),
\end{eqnarray}
and $\chi$ is the characteristic function, and similar for $\psi$, then we get
\begin{align}
\phi_<(t,z)&=\int_{\mathbb R} \frac {dk}{4\pi \omega(k)} \left[ c(k) e^{-i\omega(k)t+ikz}+c(k)^* e^{i\omega(k)t-ikz} \right],\\
\phi_\geq(t,z)&=\sum_{a=\pm}\int_{\mathbb R} \frac {dk}{4\pi \omega_a(k)n(\omega_a(k))} \left[ b_a(k) e^{-i\omega_a(k)t+ikz}+b_a(k)^* e^{i\omega_a(k)t-ikz} \right],\\
\psi_<(t,z)&=0, \\
\psi_\geq(t,z)&=\sum_{a=\pm}\int_{\mathbb R} \frac {dk}{4\pi \omega_a(k)n(\omega_a(k))} \frac {\omega^2_a(k)-k^2}{ig\omega_a(k)} \left[ b_a(k) e^{-i\omega_a(k)t+ikz}-b_a(k)^* e^{i\omega_a(k)t-ikz} \right],
\end{align}
where
\begin{align}
\omega(k)&=|k|,\\
\omega_\pm(k)&=\frac 12 \sqrt{g^2+(\omega_0+|k|)^2}\pm \frac 12 \sqrt{g^2+(\omega_0-|k|)^2},\\
n(\omega)&=1+\frac {g^2\omega_0^2}{(\omega_0^2-\omega^2)^2}.
\end{align}
In particular, $\omega_\pm$ are the positive branches corresponding to the dispersion relation
\begin{eqnarray}
k_0^2=k^2+\frac {g^2k_0^2}{k_0^2-\omega_0^2},
\end{eqnarray}
where $k_0$ is the time component of the two-momentum. \\
The coefficients $c, b_\pm$ are related by imposing the boundary conditions for $\phi$ on $z=0$. Equivalently, we can look for a basis of plane wave solutions, say, a scattering basis. After some tedious algebra 
we get the positive frequency ``basis'' $(\phi_k,\psi_k)^t$ defined by
\begin{align}
e^{i|k|t}\phi_k(t,z)=&\left[\theta(k) \left( e^{ikz}+\frac {k-q}{k+q}e^{-ikz} \right)+\theta(-k) \frac {2q}{k+q} e^{ikz}\right]\chi_{(-\infty,0)}(z)\cr
& +\left[\theta(k)  \frac {2k}{k+q} e^{iqz} +\theta(-k)\left(e^{iqz}+\frac {q-k}{k+q}e^{-iqz} \right)\right]\chi_{[0,\infty)}(z),\\
e^{i|k|t}\psi_k(t,z)&=\frac {k^2-q^2}{ig|k|}\left[\theta(k)  \frac {2k}{k+q} e^{iqz} +\theta(-k)\left(e^{iqz}+\frac {q-k}{k+q}e^{-iqz} \right)\right]\chi_{[0,\infty)}(z),
\end{align} 
where 
\begin{eqnarray}
q=q(k)=k\sqrt{\frac {k^2-g^2-\omega_0^2}{k^2-\omega_0^2}}
\end{eqnarray}
is such that $\omega(k)=\omega_a(q(k))$.\\
Notice that $q(k)$, and then the scattering basis, is defined only for $|k|<\omega_0$ or $|k|>\sqrt{\omega_0^2+g^2}\equiv \bar \omega$. This leads to a ill definiteness of the problem, which we will now investigate. 

\section{The spectral boundary conditions}
In order to understand the ill definiteness of the problem let us first discuss the simple origin of the trouble. The point is that the modes with $\omega_0\leq |k|\leq \bar\omega$ correspond to the gap in the dispersion relations in the medium 
(see the figure). 

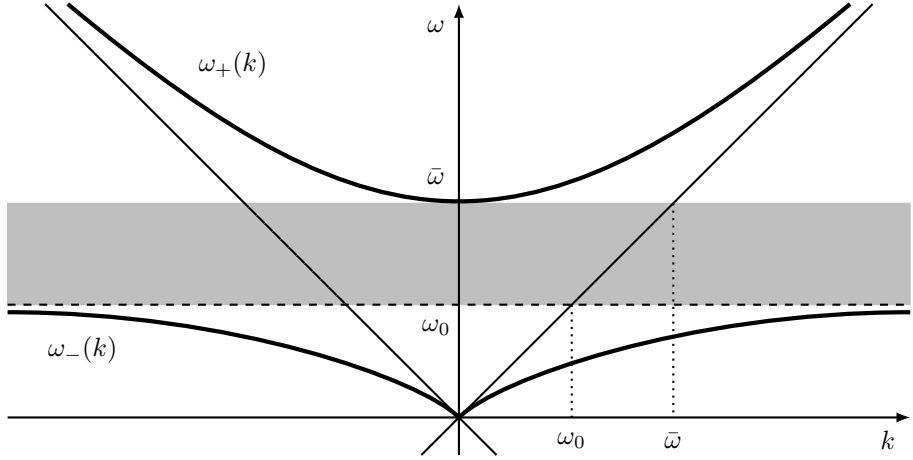
\begin{figure}[!htbp]
\begin{center}
\begin{tikzpicture}[>=latex] 
\draw [lightgray, fill=lightgray](-1,2.85) -- (11,2.85) -- (11,1.5) -- (-1,1.5) -- cycle;
\draw [thick,->](-1,0) -- (11,0);
\draw [thick,->](5,-0.5) -- (5,5.5);
\draw [thick,-,dashed](-1,1.5) -- (11,1.5);
\draw [thick,-](4.5,-0.5) -- (10.5,5.5);	
\draw [thick,-](-0.5,5.5) -- (5.5,-0.5);
\draw [thick,dotted] (6.5,1.5) -- (6.5,0);
\draw [thick,dotted] (7.85,2.85) -- (7.85,0);
\draw[ultra thick] (-0.2,5.5) .. controls (4,2) and (6,2) .. (10.2,5.5);
\draw[ultra thick] (5,0) .. controls (5.5,0.5) and (8,1.4) .. (11,1.4);
\draw[ultra thick] (5,0) .. controls (4.5,0.5) and (2,1.4) .. (-1,1.4);
\node at (4.7,1.2) {$\omega_0$};
\node at (4.7,5.2) {$\omega$};
\node at (4.7,3.2) {$\bar \omega$};
\node at (10.7,-0.3) {$k$};
\node at (6.5,-0.3) {$\omega_0$};
\node at (7.85,-0.3) {$\bar\omega$};
\node at (2,4.7){$\omega_+(k)$};
\node at (0,0.9){$\omega_-(k)$};
\end{tikzpicture}
\caption{The Sellmeier dispersion relation in the lab. The shaded region evidences the gap.}\label{figura dispersion}
\end{center}
\end{figure}

Thus, for such modes the relation $\omega(k)=\omega_a(q(k))$ cannot be satisfied. For these, $b_a=0$ and the Neuman condition on $z=0$ implies that $c(k)=0$ for the modes in the gap. From the physical point
of view this means that modes with $k$-vector in the gap cannot propagate in $z<0$, which sounds absurd! As to say that a $\phi$-laser with frequency centred, say, at $\omega=(\omega_0+\bar\omega)/2$ cannot work because somewhere
far away (no matter how much) is present a dielectric with a gap in the spectrum.\\
From the mathematical point of view, this corresponds to an incompleteness of the scattering basis, because it does not allow for describing all possible finite energy initial states, since initial states living in vacuum with modes in the gap
are complementary to the scattering basis, as we will now argue.
\subsection{Incompleteness of the scattering basis}
In some sense we can say that the scattering basis is complete in the right side, inside the matter. Indeed, if we define $k_\pm=\omega_\pm(p) p/|p|$, for $z>0$ we can write 
\begin{align}
e^{ipz} \theta(z)=&\theta(z)\theta(-p) \beta(p) \left( \frac {p+k_+}{p-k_+} \phi_{k_+}(0,z) - \frac {p+k_-}{p-k_-} \phi_{k_-}(0,z) \right)\cr
&+\theta(z)\theta(p)\beta(p) \left( \frac {2p}{p-k_+}\phi_{k_+}(0,z) -\frac {2p}{p-k_-}\phi_{k_-}(0,z) +\phi_{-k_+}(0,z)-\phi_{-k_-}(0,z) \right)\cr
&=:\theta(z) \bar \phi_p(z),
\end{align}
where
\begin{eqnarray}
\beta(p)=\left( \frac {p+k_+}{p-k_+}-\frac {p+k_-}{p-k_-} \right)^{-1}.
\end{eqnarray}
This way, the generators $e^{ipz}$ is realised for $z>0$. Notice that for $z<0$
\begin{eqnarray}
\bar\phi_p(z)=\beta(p) \left( \frac {2p}{p-k_+} e^{ik_+z}+\frac {2p}{p-k_-} e^{ik_-z}\right).
\end{eqnarray}
In a similar way we can reproduce the combination reproducing $e^{-ipz}$ in $z>0$. This does not provide an equivalent set of functions since there remain further possible combination, which are
\begin{align}
 \frac p{k_+} \phi_{k_+}-\phi_{k_-}-\phi_{-k_-}-\phi_{-k_+}&=2\theta(-z)\left( \frac p{k_+} \sin (k_+z) -i\cos (k_-z) \right),\\
 \phi_{k_+} -\frac p{k_-} \phi_{k_-}+\phi_{-k_-}+\phi_{-k_+}&=2\theta(-z)\left( \cos (k_+z) -i\frac p{k_-} \sin (k_-z) \right),
\end{align}
which do not provide a complete set of solutions since exactly the modes with vector $k$ in the gap are absent. We have looked at the field $\phi$ only, since for $\psi$ we can add arbitrary $k$ modes with frequency $\omega_0$ so that
there are no problems of completeness.\\ 
Thus we see that in order to have a complete set of solutions we should add those modes which vanish in $z>0$ and have the spectral parameter $k$ in the gap $\omega_0\leq |k| \leq \bar\omega$.\\

The completion of the basis is obtained by adding the gap modes
\begin{eqnarray}
\phi_{g,k}(z)=c(k) \sin (kz) \theta(-z).
\end{eqnarray}
Now we can construct the projection operators in order to specify the spectral boundary conditions.

\subsection{Inner product, Hamiltonian and boundary conditions}

Let us define 
\begin{eqnarray}
&&\chimag=\chi_{[0,\infty)}(z),\cr
&&\chimin=\chi_{(-\infty,0)}(z).
\end{eqnarray}
We also define the symplectic matrix 
\begin{eqnarray}
\Omega:=\left[
\begin{array}{cc}
0_{2\times 2}  & -i {I}_{2\times 2} \cr
i {I}_{2\times 2} & 0_{2\times 2}
\end{array}
\right],
\label{symma}
\end{eqnarray}
the multicomponent field 
\beq
\Psi :=\left(
\begin{array}{c}
\phi\cr
\psi\cr
\pi_\phi\cr
\pi_\psi
\end{array}
\right),
\eeq
and also the inner product 
\beq
<\Psi_1,\Psi_2> =(\Psi_1,\Omega \Psi_2),
\eeq
where $(.,.)$ stays for the standard product in $L^2$. 
Then we obtain 
\beq
<\Psi_1,\Psi_2> =-\frac{i}{2} \int dz \left[ \phi_1^\ast \overleftrightarrow{\partial}_0 \phi_2+(\psi_1^\ast  \overleftrightarrow{\partial}_0 \psi_2+
g (\psi_1^\ast \phi_2 - \phi^\ast_1 \psi_2) )\chimag \right].
\eeq
In order to better understand the problem of completeness and of the boundary conditions, we introduce the Hamiltonian operator $H$ such that 
the equations of motion are written in a Hamiltonian form:
\beq
\partial_0 \left(
\begin{array}{c}
\phi\cr
\psi\cr
\pi_\phi \chimag\cr
\pi_\psi \chimag
\end{array}
\right)
= \left(
\begin{array}{c}
\pi_\phi\cr
(\pi_\psi+g \phi)\chimag\cr
\partial_z^2\phi  -g \pi_\psi \chimag -g^2 \phi \chimag\cr
-\omega_0^2 \psi \chimag
\end{array}
\right) 
\Longleftrightarrow \partial_0 \Psi = H \psi,
\eeq
where
\beq
H:= \left[
\begin{array}{cccc}
0 & 0 & 1 & 0\cr
g \chimag & 0 & 0 &\chimag\cr
\partial_z^2 -g^2 \chimag  & 0 & 0 &-g \chimag\cr
0 & -\omega_0^2 \chimag & 0 & 0
\end{array}
\right].
\eeq
It is not difficult to show that the operator 
\beq
\hat{H} := i H
\eeq
is formally selfadjoint with respect to the inner product $<.,.>$ defined above. In order to verify this, we must integrate by part the 
derivative contributions appearing in the operator $\hat H$. So doing, we discover that in $z=0$ some boundary terms appear, which 
are a priori possible hindrances to the hermicity of the operator itself. These boundary terms are of the form
\beq
\phi (0^+) \partial_z \phi (0^+)-\phi (0^-) \partial_z \phi (0^-);
\eeq
we can get rid of them in three ways: a) we can impose the continuity of the field and of its derivative at $z=0$, as in the case of a standard 
problem of scattering in presence of a step-like potential barrier; b) we can impose Dirichlet boundary conditions at $z=0$; c) we can apply Neumann 
boundary conditions at $z=0$. 
We point out that our function space, endowed with the aforementioned inner 
product, is a Krein space (negative norm states, which amount to antiparticles in a quantum field theory framework, appear). Selfadjointness 
means in this case that 
\beq
H_c :=(\Omega \hat{H} \Omega)^\dagger 
\eeq
coincides with $\hat{H}$. We note that the spectrum of the Hamiltonian operator coincides with the one-particle frequencies $\omega,\omega_\alpha$ 
which were discussed in the previous section. Note also that such frequencies are conserved, as separation of variables easily shows.\\
In order to judge about the selfadjointness problem, we could proceed as follows: let us consider eigenstates $\Psi = \exp (-i \omega t) f(z)$, with  
the spatial part $f(z)$ smooth with compact support.
This requirement is such that boundary terms immediately disappear, as in the case (a) above, but there remains a problem. 
Indeed, such a choice of functional space implies that the fields and their partial derivatives with respect to $z$ are continuous in $z=0$, and this requirement 
eliminates the boundary terms. Still, there is an unsatisfactory property from a physical point of view, i.e. the electric field would vanish for all the 
frequencies belonging to the mass gap, which is not a physically acceptable property for what was discussed previously. Then we must provide a further 
specification for the physical domain of $\hat{H}$. If we require that Dirichlet boundary conditions are satisfied at $z=0$ for all the frequencies in the mass gap of our dispersion law, then we obtain a satisfactory behaviour for our operator.\\ 
As it is a selfadjoint operator, its eigenfunctions are orthogonal for $\omega \not = \omega'$. We can define a projector $P_{gap}$ which is such that 
the boundary conditions become
\beq
B \phi := P_{gap} \phi|_{z=0}=0.
\eeq
With these boundary conditions one takes into account properly the fact that, in the mass gap, $z=0$ becomes a sort of infinite barrier which expels 
the field from inside the dielectric medium because the Sellmeier displays a mass gap region in the spectrum. This unusual feature is due to the 
transparency of the medium, which simplifies greatly the quantum version of the model but does not allow to gain a completely satisfactory model. 
Notice that continuity of the field and of its $z$-derivative is not a boundary condition, but takes into account the finiteness of the barrier for all frequencies 
outside the mass gap (the electric field can live both inside and outside). Our analysis for the Hopfield model is in agreement with the short discussion appearing in 
\cite{maradudin} and concerning the exciton behaviour in the case of absence of spatial dispersion  (see section II therein).\\
It is also remarkable that a spectral boundary condition of the type
\beq
B_s \phi := P_{gap} \phi|_{z=0}\oplus \partial_z (1-P_{gap}) \phi|_{z=0}=0,
\eeq
would produce unphysical results, as it would imply no transmission in the scattering basis for the field $\phi$.\\

\section{Solitonic solution}
In this section, we introduce a nonlinear term in the polarization field $\psi$ and look for solitonic solutions of the field equations. 
We recall that solitons in Kerr dielectric media are usually derived in the framework of the so-called nonlinear Schroedinger equation 
(NLS). See e.g. \cite{ablowitz} for NLS, and \cite{agrawal} for solitonic solutions in fiber optics. See also \cite{drummond-book} for further 
discussion.\\
We look for a static solution in the dielectric, in the comoving frame; we rewrite the hamiltonian action~\eqref{action1} in a covariant form, and add a self interaction $\psi^4$ term simulating the Kerr effect, i.e. generating a dielectric perturbation 
moving with substantially constant velocity in the bulk dielectric medium:
\beq
\label{action2}
S[\phi,\psi] = \int_{\mathbb R} dt \Biggl\{ \int_{\mathbb R} \frac{1}{2} \partial_\mu \phi \partial^\mu \phi \, dz +
\int_{z\geq 0} \biggl[ \frac{1}{2} (v^\mu \partial_\mu \psi)^2 - \frac{{\omega_0}^2}{2} \psi^2 -
g \phi v^\mu \partial_\mu \psi - \frac{\lambda}{4!} \psi^4 \, dz \biggr] \Biggr\} .
\eeq
Note that~\eqref{action1} is obtained by means of $v^\mu=(1,0)$ and $\lambda=0$. \\
We set $v=\gamma V$ for the velocity of the dielectric perturbation on the $z$-axis, and imposing the staticity $\phi(t,z) \equiv \phi(z)$,
$\psi(t,z) \equiv \psi(z)$, we obtain
\begin{align}
\psi(z) & = \frac{a}{\cosh(bz)} , \\
\phi(z) & = \frac{2agv}{b} \arctan \biggl[ \tanh \biggl(\frac{b}{2}z\biggr) \biggr] ,
\end{align}
where
\begin{align}
a & := \sqrt{ \frac{12}{\lambda} (g^2v^2 - \omega_0^2) } , \\
\label{b}
b & := \frac{1}{v} \sqrt{g^2v^2 - \omega_0^2 } .
\end{align}
Note that the static solution exists if and only if $b>0$, so $v>\omega_0/g$. \\
All of this is true in the comoving frame; in order to obtain solutions in the lab frame, we boost our functions by
$z \to \gamma(z-Vt)$, so
\begin{align}
\label{psidiel}
\psi(t,z) & = \frac{a}{\cosh(b\gamma(z-Vt))} , \\
\label{phidiel}
\phi(t,z) & = \frac{2agv}{b} \arctan \biggl[ \tanh \biggl(\frac{b}{2}\gamma(z-Vt)\biggr) \biggr] .
\end{align}
This is our solution in dielectric ($z>0$). 
It is useful to provide also for $z>0$ 
\beq
\phi' (t,z)  = agv \gamma \frac{1}{\cosh (b \gamma(z-Vt))}.
\eeq
Now, we want to glue this solution with that in the vacuum; since dielectric is moving, the field in vacuum is time-dependent, so
the general solution to $\square\phi = 0$ is
\beq
\phi(t,z) = \alpha(z-t) + \beta(z+t) ,
\eeq
while $\psi=0$ in $z<0$. \\
We have to impose the gluing conditions at $z=0$ for every time $t$: we want the continuity of the filed $\phi$ and of its normal derivative
$\partial_n\phi$ (that is equivalent, in the lab frame, to $\partial_z\phi$). \\
We obtain
\begin{align}
\alpha'(z-t) & = \frac{a\gamma gv(1+V)}{2\cosh(b\gamma V(z-t))} , \\
\beta'(z+t) & = \frac{a\gamma gv(1-V)}{2\cosh(b\gamma V(z+t))} .
\end{align}
We can interpret our solution in this way: we have a progressive wave $\alpha(z-t)$ that clashes the dielectric.
After the collision, we will have a reflected wave $\beta(z+t)$ and a transmitted wave~\eqref{phidiel}, plus the polarization field~\eqref{psidiel}. \\
In order to confirm our physical interpretation, we are going to calculate the total energy of the system.

\subsection{Solitonic energy}
Hamiltonian of the theory (in lab frame) is
\beq
\label{hamilton}
\mathcal{H} = \frac{1}{2} \bigl(\dot{\phi}^2 + {\phi'}^2 + \dot{\psi}^2 + \omega_0^2 \psi^2\bigr) + \frac{\lambda}{4!} \psi^4 ;
\eeq
we split our calculus in vacumm and dielectric parts: we start with the vacuum part. \\
In vacuum, hamiltonian is reduced to ($\psi=0$)
\beq
\mathcal{H}_V = \frac{1}{2} \bigl(\dot{\phi}^2 + {\phi'}^2\bigr) =
\frac{a^2\gamma^2g^2v^2}{4} \biggl[ \frac{(1+V)^2}{\cosh^2(b\gamma V(z-t))} + \frac{(1-V)^2}{\cosh^2(b\gamma V(z+t))} \biggr] ;
\eeq
so, the total energy in $z<0$ is
\beq
E_V = \int_{-\infty}^0 dz \, \mathcal{H}_V = \frac{a^2\gamma g^2v^2}{2bV} \bigl[ (1+V^2) - 2V \tanh(b\gamma Vt) \bigr] .
\eeq
We note immediatly that $E_V$ is strictly positive, and that is a decrescent function of $t$, as we expected. \\
Hamiltonian in dielectric is equivalent to~\eqref{hamilton}, and explicitly
\beq
\mathcal{H}_D = \frac{a^2\gamma^2g^2v^2}{\cosh^2(b\gamma(z-Vt))} ;
\eeq
energy in dielectric is
\beq
E_D = \int_0^{+\infty} dz \, \mathcal{H}_D = \frac{a^2\gamma g^2v^2}{b} \bigl[ 1 + \tanh(b\gamma Vt) \bigr] .
\eeq
$E_D$ is an increasing function, and it is positive too. \\
This calculus confirm our physical picture: we have an energy flux from $z\to-\infty$ to $z\to+\infty$, because energy in vacuum decreases
in time, while the dielectric is progressive filled and its energy increases. \\
It is worthwhile noting that energy in vacuum does not converge to zero, because there exists a reflected wave $\beta(z+t)$.
At $t\to+\infty$, both $E_V$ and $E_D$ become constant, and this situation corresponds to a ``fullfilled'' system. \\
Now, we obtain the total amount of energy: since action~\eqref{action2} is invariant under temporal translation,
total energy is a Noether charge, so we expect that it is time-indipendent. Indeed
\beq
E_{tot} = E_V + E_D = \frac{a^2\gamma g^2v^2}{2bV} (1+V)^2 ;
\eeq
as expected it is time-indipendent, and $E_{tot} > 0$.\\
It is interesting to observe that our solitonic solution fulfills four properties which are associated as `definitory properties' to solitons  in 
\cite{marmo}: 1) finite total energy; 2) finite, non-singular and localized energy density, where localization means that at any time $t$ 
the region where $\mathcal{H}\geq \delta$, for any $\delta \in (0,\max_z \mathcal{H})$, is bounded; 3) the solitonic solution is non-singular; 
4) the solitonic solution is non-dissipative (in the sense that $\max_z \mathcal{H}$ does not vanishes as $t\to \infty$). There is also a fifth 
(and last) property, i.e. classical stability in the sense of Lyapunov, which is shown to be implemented in the following section.

\subsection{Transmission and reflection coefficients}

The naive expectation for the process at hand is the following: one would expect that the scattering of the 
solitonic wave starts from the left $z\ll 0$  at very early times $t\to -\infty$, with a progressive wave moving towards 
the interface $z=0$, and that at very far times in the future $t\to +\infty$ one gets two separate packets, the first 
counter-propagating in the vacuum region $z\ll 0$ (reflected wave) and the second one progressing in the dielectric 
medium $z\gg 0$ with velocity $V$ (transmitted wave) . We show that this is the case, and provide the reflection and the transmission 
coefficients.\\

Usually, one could approach the problem by using the component $J_z$ of the current density. In the present case, 
this is not possible because there is no charge displacement in the medium and then $J_z=0$. We can then 
approach the problem by recalling that the canonical stress energy tensor can be calculated, and in particular the 
component $T_{0z}$ represents the flux of energy through a surface orthogonal to the $z$-direction. To be more specific, 
we calculate $T_{0z}$ for the only field which propagates in our system: the `electromagnetic field' $\phi$.\footnote{ Indeed, 
there is no real propagation of the polarization field $\psi$, which is present only in the $z\geq 0$ region and is 
in some sense pathological: it represents fixed dipoles oscillating around a fixed position in space, with a given 
frequency $\omega_0$, and exists just in the dielectric medium. Then, it cannot be involved in energy transport. 
It can be noted that its contribution to the stress-energy tensor would be non-symmetric, even if it is just a scalar field. 
A Belinfante-Rosenfeld procedure could be taken into account, or even a Abraham-like tensor could be set up for the 
polarization part. Still, a scattering picture would be meaningless, as, by definition, $\psi$ is just present for $z\geq 0$.}
We get 
\beq
T^{\mu \nu}=\frac{\partial L}{\partial \partial_\mu \phi} \partial^{\nu} \phi-\eta^{\mu \nu} L,
\eeq
and, in particular, we are interested in 
\beq
T^{0z}=(\partial^0 \phi)(\partial^z \phi) =:S^z,
\eeq
where $S^z$ is the equivalent of the Poynting vector for the electromagnetic field. Then we get 
\beq
S_z=-\dot{\phi} \phi'.
\eeq
Let us consider the vacuum region $z<0$: 
\beq
S_z = (\alpha')^2-(\beta')^2.
\eeq 
In the dielectric region $z\geq 0$ we obtain
\beq
S_z= V (\phi_D')^2,
\eeq
where we indicate with $\phi_D$ the field $\phi$ in the dielectric region. 
Let us consider the field at $t\ll 0$ and for $z \ll 0$. In order to get a field which is 
appreciably different from zero, we should impose $t \sim z$, in such a way to 
obtain the peak value for $\alpha'$, whereas $\beta' \sim 0$ and $\phi_D'= 0$. 
In such a situation, we have
\beq
\alpha' (z-t) \sim \frac{a \gamma g v (1+V)}{2}.
\eeq
This situation represents the initial pulse which moves towards $z=0$ from the 
left. It is also useful to note that for $t\ll 0$ and $z>0$ one has $\phi_D' (z+V |t|)\sim 0$.\\
Let us now consider $t\gg 0$, i.e. the final state. In the region $z\ll 0$, we get 
a field appreciably different from zero only for $t \sim -z$, i.e. we get a 
reflected packet with 
\beq
\beta' (z+t) \sim \frac{a \gamma g v (1-V)}{2}.
\eeq
In the dielectric region, we obtain a non-vanishing contribution only for 
$t\sim z/V$, which correspond to the peak of the transmitted packet. 
Summarizing, we have the Poynting vector for the initial packet, for the reflected one 
and for the transmitted one respectively:
\beqn
S_z (t\ll 0,z\ll 0, z\sim t) &\simeq& (\alpha')^2 = \frac{a^2 \gamma^2 g^2 v^2 (1+V)^2}{4},\\
S_z (t\gg 0,z\ll 0, z\sim -t) &\simeq& (\beta')^2 = \frac{a^2 \gamma^2 g^2 v^2 (1-V)^2}{4},\\
S_z (t\gg 0,z\gg 0, z\sim V t) &\simeq& V(\phi_D')^2.
\eeqn
We can then obtain the reflection and the transmission coefficients:
\beqn
R: &=& \frac{S_z (t\gg 0,z\ll 0, z\sim -t) }{S_z (t\ll 0,z\ll 0, z\sim t)}=\left( \frac{1-V}{1+V} \right)^2,\\
T:  &=& \frac{S_z (t\gg 0,z\gg 0, z\sim V t) }{S_z (t\ll 0,z\ll 0, z\sim t)}= \frac{4 V}{(1+V)^2}.
\eeqn
They satisfy
\beq
R+T=1.
\eeq
It is interesting to point out that the Poynting vector $S_z$ is continuous at the surface $z=0$, as it would be 
expected for the Poynting vector in the full electromagnetic case: 
\beq
(\alpha')^2 (t) - (\beta')^2 (t)= V (\phi_D')^2 (t).
\eeq
Indeed, one obtains
\beq
\frac{a^2 \gamma^2 g^2 v^2 (1+V)^2}{4 \cosh^2 (b \gamma V t)}-
\frac{a^2 \gamma^2 g^2 v^2 (1-V)^2}{4 \cosh^2 (b \gamma V t)}=
\frac{a^2 \gamma^2 g^2 v^2 V}{\cosh^2 (b \gamma V t)}.
\eeq

\section{Stability}

Stability of solitons is a nontrivial problem, which has to face in dimension greater than two with a strong 
no-go theorem due to Hobart and Derrick. A subtle distinction between absolute stability and stability in the 
sense of Lyapunov has to be taken into account, as discussed e.g. in \cite{skyrme-book} and in 
\cite{jackiw-rossi}. We consider for simplicity the case of infinite dielectric, but extension to our previous framework 
are possible (see below). In our analysis, we follow the ideas contained in \cite{jackiw-rossi}, and 
we are able to infer stability of our soliton solution.\\
The starting point consists in writing the Hamiltonian operator as a function of the field momenta and the fields 
themselves: 
\beq
H=H(\pi_\phi,\pi_\psi,\phi,\psi).
\eeq
Then one has to consider an expansion of $H$ up to the second order in the field and momenta 
variations $\delta \pi_\phi, \delta \pi_\psi,\delta \phi, \delta \psi$ around the soliton solution, taking into account that the first order 
contribution vanishes (as the soliton solution is `on shell', i.e. satisfies the Hamiltonian equations of motion). By defining 
\beqn
\delta P &=& 
\left(
\begin{array}{c}
\delta \pi_\phi\\
\delta \pi_\psi\\
\end{array}
\right),\\
\delta Q &=& 
\left(
\begin{array}{c}
\delta \phi\\
\delta \psi\\
\end{array}
\right),\\
\eeqn
and indicating by $\delta P^t,\delta Q^t$ the transposed vectors, 
we get 
\beq
H(\pi_\phi,\pi_\psi,\phi,\psi) = H_0+ \frac{1}{2} \delta P^t\ T\ \delta P+ 
\frac{1}{2} \delta Q^t\ V\ \delta Q + \delta P\ G\ \delta Q+\cdots,
\eeq 
where $H_0$ is the zeroth-order contribution associated with 
the solitonic solutions, and where higher order contributions are neglected. For simplicity, we work in the lab frame. Explicitly, we get 
\beq
T=
\left[
\begin{array}{cc}
1 & 0\\
0 & 1\\
\end{array}
\right],
\eeq
\beq
G=
\left[
\begin{array}{cc}
0 & 0\\
g & 0\\
\end{array}
\right],
\eeq
and
\beq
V=
\left[
\begin{array}{cc}
-\partial_z^2 +g^2& 0\\
0 & \omega_0^2+\frac{\lambda}{2} \psi_0^2\\
\end{array}
\right],
\eeq
where $\psi_0$ corresponds to the solitonic solution. 
Then 
one finds a second order operator 
\beq
K=
\left[
\begin{array}{cc}
T & G\\
G^t & V\\
\end{array}
\right],
\eeq
where $G^t$ stays for the transposed matrix. It can be noted that, being $G\not =0$, 
in our model appear gyroscopic-like contributions, which require that stability in the Lyapunov 
sense is shifted from the requirement of minimality of the energy functional, to more general 
conditions \cite{jackiw-rossi}. 
According to the analysis in 
\cite{jackiw-rossi}, by introducing the symplectic matrix (\ref{symma}), a stationary solution 
\beq
e^{-i \omega t} X(z)
\eeq
where $X(z)$ is the suitable vector function depending only on $x$, is stable in the sense of Lyapunov 
if the equation 
\beq
\det (K-\omega \Omega) =0
\eeq
admits only real solutions $\omega\in {\mathbf{R}}$. Note that the symplectic eigenvector $X$ 
satisfies 
\beq
K X = \omega \Omega X.
\eeq
In the present case, by using with some ingenuity the rule 
\beq
\det A = \det A_{11} \det (A_{22} -A_{21} A_{11}^{-1} A_{12}),
\eeq
which holds for a block square matrix $A$ with equal square blocks $A_{11},A_{12},A_{21},A_{22}$
\beq
A=
\left[
\begin{array}{cc}
A_{11} & A_{12}\\
A_{21} & A_{22}\\
\end{array}
\right],
\eeq
one obtains 
\beq
\det (K-\omega \Omega) = \det 
\left[
\begin{array}{cc}
-\partial_z^2 -\omega^2 & -i\omega g\\
i\omega g & \omega_0^2 -\omega^2+\frac{\lambda}{2} \psi_0^2 \\
\end{array}
\right]=0.
\eeq
We can observe that, momentarily neglecting the contribution associated with $\psi_0$, one obtains 
substantially the dispersion relation in the dielectric medium, as it is easy to realize by means of a Fourier analysis 
of the above operator. We know that $\omega$ is real when $\omega^2$ is outside the forbidden interval $(\omega_0^2,\omega_0^2+g^2)$, 
which corresponds to the well-known mass-gap in the Sellmeier dispersion relation delimiting the reality of the 
refractive index in optics. We have also to keep into account that the $\psi_0$ contribution, for positive $\lambda$, 
is a positive and bounded operator which represents a small perturbation with respect to the (unbounded) operator 
$\omega_0^2 -\omega^2$. Such  a perturbation is substantially not able to perturb in any sensible way the reality of 
$\omega$, in the sense that if we define 
\beq
s^2 :=\omega_0^2 +\frac{\lambda}{2} \sup(\psi_0^2),
\eeq
then stability in the above sense is ensured as far as 
\beqn
\omega^2 &<& \omega_0^2,\\
\omega^2 &>& s^2+g^2,
\eeqn
which correspond to the stability conditions for the case at hand.\\

In the case of semi-infinite dielectric medium, one has that the Hamiltonian $H$ is split into two expressions, 
one for the vacuum region and the other for the dielectric region. As a consequence, the operator $K$ above is split 
into two expressions too: 
\beq
K =
\left[
\begin{array}{c}
K_< \quad \quad z<0,\\
K_> \quad \quad z\geq 0,
\end{array}
\right.
\eeq
where $K_>$ is formally the same as in the infinite dielectric medium discussed above, and 
$K_<$ is a free field contribution associated only with $\pi_\phi,\phi$ (as the polarization field 
contribution vanishes). The latter contribution does not affect stability properties discussed above.

\section{Conclusions}

In the framework of  a 1+1 dimensional toy-model which consists of an half-line filling dielectric medium, simulating the electromagnetic field and the polarization field by means of two coupled scalar fields $\phi$,$\psi$ respectively, in a Hopfield-like model, we have achieved substantially 
two relevant results. The first one is that boundary conditions associated with the spectrum of quasi-particles (polaritons) 
are necessary in order to match meaningful physics. Indeed, in order to avoid an unphysical behaviour in the case of the `mass gap', one is forced to 
impose reflection at $z=0$ for the electromagnetic field $\phi_\omega$ when $\omega$ falls in the mass gap. This is a somewhat unexpected 
condition, which can be related to the requirement of transparency of the dielectric medium.
The second result is 
that, in a nonlinear model, solitonic solutions exist, whose behavior has been studied. All relevant properties of solitons are shown to be fulfilled.
Indeed, we have considered the energy behaviour both in a global sense, by studying how the energy changes with time in the different parts of our setting, 
and in a local sense, by means of the analysis, based on $T_{0z}$, which represents the flux of energy through a surface orthogonal to the $z$-direction.  
We have also studied stability in the Lyapunov sense of the solitonic solutions.



\end{document}